\documentclass[%
reprint,
superscriptaddress,
citeautoscript,
 amsmath,amssymb,
 aps,
prb,
floatfix,
]{revtex4-2}

\usepackage{graphicx}
\usepackage{subfigure}
\usepackage{siunitx}
\usepackage{dcolumn}
\usepackage{multirow}
\usepackage{blindtext}
\usepackage{bm}
\usepackage{xcolor}
\usepackage{booktabs}
\usepackage[mathlines]{lineno}

\usepackage{float}
\usepackage[hidelinks,urlcolor=blue]{hyperref}

\newcommand{\etal}{{\em{et\,al.\,}}}

\begin{document}


\title{Band-gap reduction and band alignments of dilute bismide III--V alloys}

\author{Abdul Saboor}
\affiliation{Department of Physics and Astronomy, University of Delaware, DE 19716, USA}
\affiliation{Department of Materials Sciences \& Engineering, University of Delaware, DE 19716, USA}

\author{Shoaib Khalid}
\affiliation{Princeton Plasma Physics Laboratory, P.O. Box 451, Princeton, New Jersey 08543, USA}

\author{Anderson Janotti}%
\affiliation{Department of Materials Sciences \& Engineering, University of Delaware, DE 19716, USA}


\date{\today}
\begin{abstract}

Adding a few atomic percent of Bi to III--V semiconductors leads to significant changes in their electronic structure and optical properties.  Bismuth substitution on the pnictogen site leads to a large increase in spin-orbit splitting $\Delta_{\rm SO}$ at the top of the valence band ($\Gamma_{8v}-\Gamma_{7v}$) and a large reduction in the band gap, creating unique opportunities in semiconductor device applications. Quantifying these changes is key to the design and simulation of electronic and optoelectronic devices. Using hybrid functional calculations, we predict the band gap of III--Vs (III=Al, Ga, In and V=As, Sb) with low concentrations of Bi (3.125\% and 6.25\%), the effects of adding Bi on the valence- and conduction-band edges, and the band offset between these dilute alloys and their III--V parent compounds. As expected, adding Bi raises the valence-band maximum (VBM). However, contrary to previous assumptions, the conduction-band minimum (CBM) is also significantly lowered, and both effects contribute to the sizable band-gap reduction.  Changes in band gap and $\Delta_{\rm SO}$ are notably larger in the arsenides than in the antimonides. We also predict cases of band-gap inversion ($\Gamma_{6c}$ below $\Gamma_{8v}$), and $\Delta_{\rm SO}$ larger than the band gap, which are key parameters for designing topological materials and for minimizing losses due to Auger recombination in infrared lasers.

\end{abstract}

\maketitle
\section{Introduction \label{sec:in}}

Incorporation of a small amount of Bi into conventional III–V semiconductor materials leads to large band-gap reduction and a significant increase in spin-orbit splitting \cite{Janotti2002, Tixier2003, Kudrawiec2011, Wang2013, Kurdawiec2014, Polak2015}. These effects make dilute bismide III--V alloys (III--V$_{1-x}$Bi$_x$, $x<10\%$) of great interest to mid-infrared optoelectronic devices such as lasers operating in the telecom wavelength (1.3–1.5 $\mu$m) \cite{Song2013,Sweeney2013, Sweeney2014}, short-wave infrared detectors for hot objects and near-IR imaging (2.5-5$\mu$m) \cite{Glady2025,carrasco_sensitivity_2025,webster_molecular_2025}, and long-wavelength photodetectors that compete with HgCdTe (8-12 $\mu$m) for remote sensing and astronomy\cite{Wen2015}.
The increased spin-orbit splitting in Bi-containing III--V alloys also offers the potential to reduce or eliminate the non-radiative Auger recombination and inter-valence-band absorption processes that dominate the performance of near- and mid-infrared light-emitting diodes and lasers \cite{Sweeney2013,Sweeney2014,Broderick2013}. Finally, starting from the narrow band gap III--V semiconductors, such as InAs and InSb, it could be possible to reduce to band gap enough to invert it, making it negative, i.e., $\Gamma_{6c}$ below $\Gamma_{8v}$, turning the Bi containing dilute alloys into topological insulators or semimetals \cite{Suzuki2013, Huang2014}, in analogy to that demonstrated in HgCdTe/CdTe heterostructures \cite{Konig2007}, with possible applications in THz photonics \cite{Ruffenach2017}.

Fabricating thin films of III--(V, Bi) dilute alloys with high structural quality remains quite challenging. The atomic size mismatch of P, As, and even Sb vs Bi is large (Table~\ref{table1}); the bonding strength between the group-III metals (Al, Ga, and In) and Bi is much weaker compared to those with P, As, and Sb; and the vapor pressure of Bi is quite low, so that Bi atoms tend to float to the surface and evaporate even using low temperatures in the thin-film deposition \cite{Vardar2013, Ludewig2015}. These factors combine to severely limit the solubility of Bi in III--Vs \cite{Ma1991, Yoshimoto2003, Lu2009, Devenson2012}. Still, GaAs$_{1-x}$Bi$_x$ \cite{Verma2001, Oe2002, Tixier2003, Lu2008, Lu2009}, InGaAs$_{1-x}$Bi$_x$ \cite{ Feng2005,Devenson2012}, GaSb$_{1-x}$Bi$_x$ \cite{Das2012}, InAs$_{1-x}$Bi$_x$ \cite{Ma1991, Okamoto1999, Verma2001} and InSb$_{1-x}$Bi$_x$ \cite{Wagner2001} with dilute amounts of Bi have been demonstrated. Of crucial importance is how the electronic structure of these dilute alloys differ from that of the III--V parent compounds, meaning not only the band gap and the spin-orbit splitting but also the absolute position of the valence-band maximum (VBM) and the conduction-band minimum (CBM). Knowing this set of parameters is essential for many semiconductor device designs.

\begin{table}
\caption{Atomic (covalent) radius \cite{Beatriz2008} and first ionization energy (IE) \cite{NIST_ASD} of the pnictogens P, As, Sb, and Bi.  The IE values are inversely related to the valence $p$ orbital energies, i.e., 
$\epsilon [\mathrm{P}(3p)] < \epsilon [\mathrm{As}(4p)] < \epsilon [\mathrm{Sb}(5p)] < \epsilon [\mathrm{Bi}(6p)]$. }
\begin{ruledtabular}
\begin{tabular}{ccr}
Atom & Radius (\AA) & IE (eV) \\ \midrule
P  &   1.07  & 10.487  \\
As &   1.19	 &  9.788  \\
Sb &   1.39  &  8.608  \\
Bi &   1.48  &  7.285
\end{tabular}
\end{ruledtabular}
\label{table1}
\end{table}

A simple model has been proposed to explain the effects of adding small amounts of Bi to III--Vs. In analogy to the band anticrossing model (BAC) used to explain the electronic structure of dilute (In,Ga)As$_{1-x}$N$_x$ alloys, where a large band-gap reduction was observed and attributed to changes in the conduction band \cite{Shan1999}, the valence-band anti-crossing model (VBAC) has been proposed to explain the effects of Bi in III--Vs \cite{Broderick2012, Deng2010, Zide2013, Alberi2007}.  In this case, it is assumed that Bi in III--Vs only affects the top of the valence band, leaving the conduction band unchanged. The band-gap reduction is then solely attributed to a lift in the VBM caused by the interaction between the supposedly lower-lying and localized Bi 6$p$ states and the higher-lying VBM of the III--Vs.  One of the problems with this simple model is that if we consider the ionization energies (IE) of the group-V atoms (see Table~\ref{table1}), the energy of the valence $p$ orbitals follows the ordering
$\epsilon [\mathrm{P}(3p)] < \epsilon [\mathrm{As}(4p)] < \epsilon [\mathrm{Sb}(5p)] < \epsilon [\mathrm{Bi}(6p)]$,
which is consistent with the trends for the natural valence-band offsets between the III--V semiconductors where the VBM increases (ionization potential (IP) decreases) for the III--V compounds from P, As, to Sb \cite{Wei1998, VandeWalle2003, Oba2014}.  Therefore, we would expect the Bi $p$ bands to lie higher in energy than the Sb, As, and P $p$ bands in these III--V compound semiconductors.

Density functional theory (DFT) calculations have been reported for the dilute III--V$_{1-x}$Bi$_x$ alloys, showing that the band gap decreases with increasing Bi concentration, along with a large band-gap bowing \cite{Janotti2002, Deng2010, Reshak2012, Abdiche2010, Mbarki2011}.  However, most of these calculations use the standard local or semilocal approximations (LDA or GGA) for the exchange and correlation potential, leading to a severe underestimation of band gaps, and, therefore, large uncertainties in the results and conclusions are expected. Besides, these calculations do not track the changes in the VBM and CBM of the alloys with respect to those of the parent compounds, but only the band gap and the size of the spin-orbit splitting. The former quantities are also essential to understanding the effects of the Bi addition. 

Calculations based on DFT-LDA for structure optimization combined with the Tran and Blaha modified Becke Johnson LDA (TB09 MBJLDA) functional for electronic structure calculations were used to study GaP, GaAs, InAs, InP, InAs, and InSb with 3.7\% Bi\cite{Polak2015}. It was found that the conduction band is lowered with the addition of Bi, contrary to the VBAC model. However, in these TB09 MBJLDA functional calculations, the band gaps of the III-Vs were adjusted to the experimental data by modifying the scaling factor that modulates the exchange-correlation potential; and the band-alignment calculations relied on just aligning the Kohn-Sham potential in the supercell containing one Bi with that in the supercell of the pure III--V compound, without considering changes in the volume due to Bi, adding uncertainty to the results.  Therefore, it is yet to be seen how the results of Ref.~\onlinecite{Polak2015} compare with the results of more accurate but much more computationally expensive hybrid functional calculations. 

Using a hybrid functional that predicts band gaps, band alignments, and ionization potentials in much better agreement with experimental values than DFT-LDA or GGA \cite{Oba2014}, we compute changes in the band gap, variations of spin-orbit splitting with Bi concentration, and band offsets between dilute III--(V,Bi) alloys and their parent compounds.  We focus our investigations on Bi in AlAs, GaAs, InAs, AlSb, GaSb, and InSb, in concentrations of 3.125\% and 6.25\%, and do not consider phosphides due to too large a mismatch between P and Bi.  We find that adding Bi not only pushes the VBM upward but also lowers the CBM considerably, with rates that significantly differ from the previous report based on the parameterized TB09 MBJLDA functional\cite{Polak2015}. The effects in the arsenides are much larger than those in the antimonides, and this is attributed to the differences in atomic orbital energies and atomic size mismatches. The downward shift in the CBM is attributed to a volume effect, i.e., as the Bi concentration increases, the volume increases, lowering the CBM due to its antibonding character.  Finally, we predict that InAs$_{1-x}$Bi$_x$ with $\sim$10\% Bi will result in inverted bands, leading to a topological insulator phase.
 
\begin{figure*}
\centering
\includegraphics[width=0.85\textwidth]{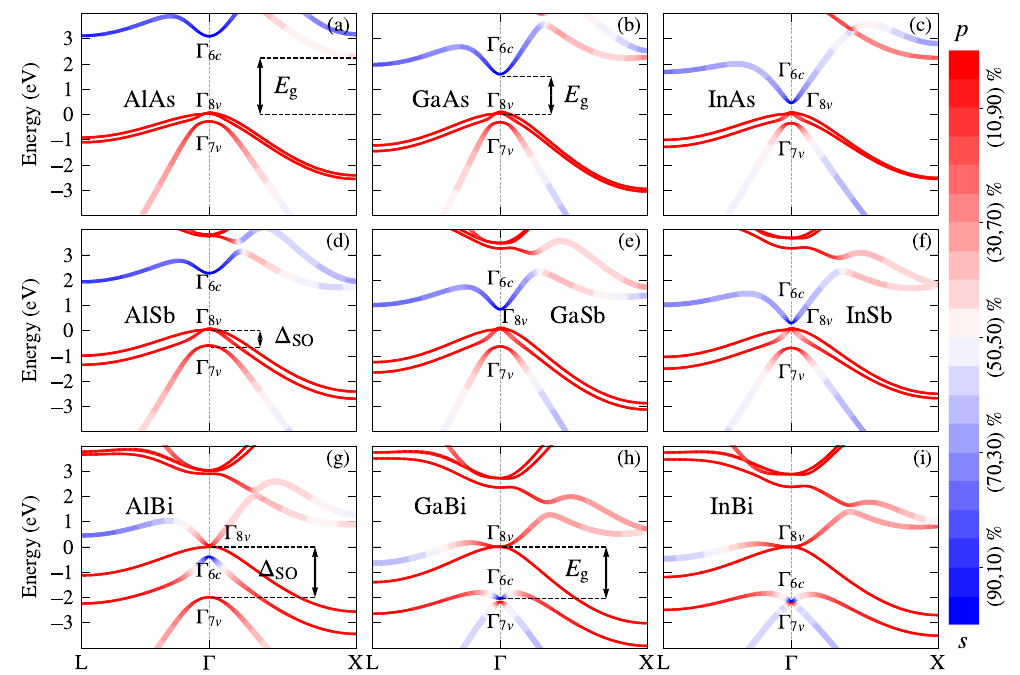}
\caption{Calculated electronic band structures of the III--V compounds, with III=Al, Ga, In and V=As and Sb, along those of AlBi, GaBi, and InBi, all in the zinc blende phase. Spin-orbit coupling is included, and the valence-band maximum (VBM) was set to zero in each case.  $\Gamma_{7v}$ is split-off band and the difference to $\Gamma_{8v}$ determines the strength of the spin-orbit coupling $\Delta_{\text{SO}}$. For the III--Bi, $\Gamma_{6c}$ is below $\Gamma_{8v}$, giving negative (inverted) band gaps. The color code represents the $s$-orbital (blue) vs $p$-orbital (red) contributions to each state, and the thickness of the lines represents the strength of the orbital contribution.}
\label{fig1}
\end{figure*}

\section{Computational Methods and Details of the Calculations \label{sec:cm}}

The calculations of dilute III--V$_{1-x}$Bi$_x$ alloys are based on the density functional theory \cite{Hohenberg1964, KohnSham1965} with the screened hybrid functional of Heyd, Scuseria, and Ernzerhof (HSE06)\cite{HSE03, HSE06}, and the projector-augmented wave (PAW) potentials \cite{Kresse1999} as implemented in the VASP code \cite{Kresse1993, Kresse1994}. Only the electrons in the highest $s$ and $p$ orbitals of the group III and V elements were considered as valence electrons; the semi-core $d$ electrons of Ga and In were treated as core electrons in the PAW potentials. The parent III--V binaries consist of combinations of the cations Al, Ga, In, and the anions As and Sb. We set the mixing parameter $\alpha$ in the HSE06 to 0.33 for AlAs and GaAs, 0.30 for InAs, and 0.27 for AlSb, GaSb and InSb for an accurate description of band gaps (low temperature values \cite{MadelungEd3}), while band gaps of AlBi,GaBi and InBi are calculated at the default value of $\alpha$=0.25 for the exact exchange fraction, since their band gaps in the zinc blende phase are unknown. Since our calculations are in the dilute Bi limit, we employed the same values of exact exchange fraction as their III--V parent compound. 

To simulate the alloys, we used supercells containing 64 atoms, with one and two anion atoms replaced with Bi corresponding to dilute III--V$_{1-x}$Bi$_x$ alloys with $x$=3.125\% (1/32) and 6.25\% (2/32) concentrations. In the case of two Bi atoms in the supercell (6.25\%), we selected configurations that maximize the Bi-Bi distance, assuming a repulsion based on strain. All the atomic positions and volume of the supercells are optimized, minimizing forces, stress tensor, and total energy. The effects of Bi clustering are not taken into account, and we expect that, in the dilute concentrations considered here and strain considerations associated with Bi substitutions resulting in an ``effective repulsion", these effects are minimized. 

For the 2-atom primitive cells of the parent III--V compounds, we used a 6$\times$6$\times$6 $k$-point mesh for the integration over the Brillouin zone, and for the 64-atom supercells we used a 2$\times$2$\times$2 $k$-point mesh, with a cutoff of 400 eV for the plane-wave basis set (convergence tests are included in the Supplemental Material \cite{SuppMat}). After all the equilibrium structures were determined, we performed calculations including spin-orbit coupling (SOC) to determine band gaps and spin-orbit splittings. 

The band-offset calculations were performed in two steps: First, we computed the VBM of the parent compound and the III--V$_{1-x}$Bi$_x$ alloy with respect to the average electrostatic potential in predetermined bulk-like regions in each material, in separate calculations. Then, we aligned these average electrostatic potentials through a third calculation using a superlattice geometry. 
We constructed 8$\times$8 superlattices of III--V/III--V$_{1-x}$Bi$_x$ along the nonpolar (110) direction, with one side representing the parent compound with 128 atoms and the other side with the III--V$_{1-x}$Bi$_x$ alloy also with 128 atoms. For the bulk-like regions in the alloys within the superlattice, we consider spheres centered on As or Sb atoms located on (110) planes far from Bi. Equivalent atoms are selected within the side of the parent compound in both the superlattice and the separate bulk calculations.  Since the VBM (in the separate III-V and alloy calculations) is referenced to the average electrostatic potential, and that depends on the volume (and specific atoms) of the cell, we constrained the volume of the alloy in the superlattice to be exactly the same as in the separate alloy calculation, imposing the in-plane lattice parameter to be that of the parent compound. In this manner, we are calculating the natural band alignment, without including the effects of epitaxial strain. Note that only the positions of the atoms within two atomic layers of the two equivalent interfaces are allowed to relax.  
\section{Results and Discussion \label{sec:rd}}

\subsection{Lattice parameters, band gap, and spin-orbit splitting of the III\texorpdfstring{$-$}{-}V parent compounds}

The calculated lattice parameter $a$, band gap $E_g$, and spin-orbit splitting $\Delta_{\rm SO}$ of AlAs, GaAs, InAs, AlSb, GaSb, and InSb, using the HSE06 hybrid functional, are listed in Table~\ref{table2}.  Lattice parameters are slightly overestimated, except for GaSb, while band gaps are slightly underestimated, except for AlSb. Spin-orbit splittings are slightly overestimated for the arsenides and slightly underestimated for the antimonides. Overall, the results are in good agreement with experimental data \cite{MadelungEd3}.

The calculated electronic band structures of the parent III--V compounds are shown in Fig.~\ref{fig1}. For the bismides AlBi, GaBi, and InBi, in the metastable zinc blende phase, we find large and negative band gaps, i.e., the $\Gamma_{6c}$ states with dominating cation contribution (Al 3$s$, Ga 4$s$, and In 5$s$, respectively) are well below $\Gamma_{8v}$, with dominating Bi 6$p$ contribution. The inverted band gaps for AlBi, GaBi, and InBi at $\Gamma$ are -0.449 eV, -2.059 eV, and -2.154 eV, while the spin-orbit splitting $\Delta_{\rm SO}$ are 2.055 eV, 2.189 eV, and 2.203 eV, respectively.  For comparison, our predicted band gap values for GaBi and InBi [$E_g(\Gamma_{6c}-\Gamma_{8v})$] are much larger in magnitude than previously reported values of -1.45 and -1.63 using DFT-LDA with added atom-dependent, spherical potential inside the muffin-tin spheres centered at each atomic and interstitial sites to correct the band gap\cite{Janotti2002}, and -1.60 and -1.89 eV using the parameterized TB09 MBJLDA functional \cite{Polak2015}. Note that the band gaps of the III--Bi do not enter into our analysis of the band gaps of the dilute III--V$_{1-x}$Bi$_x$ alloys.

\begin{figure}
\centering
\includegraphics[width=0.4\textwidth]{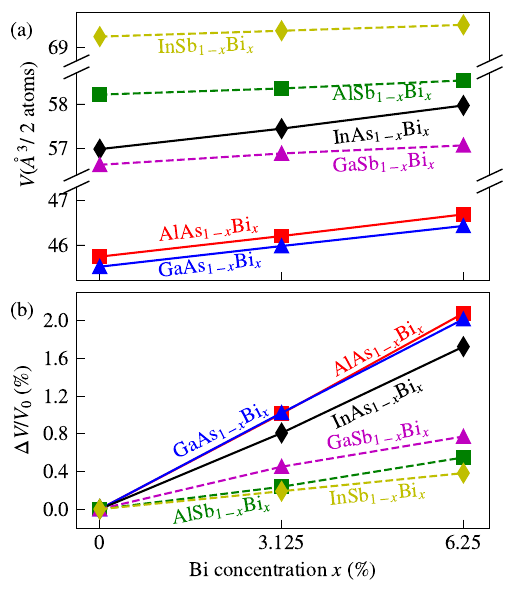}
\caption{(a) Volume (per two atoms) of dilute III--V$_{1-x}$Bi$_x$ alloys as function of Bi concentration. (b) Changes in the volume (per two atoms) of the 
dilute III--V$_{1-x}$Bi$_x$ alloys relative to the equilibrium volume of the III--V parent compound ($\Delta V/V_{0}$) as a function of Bi concentration. }
\label{fig2}
\end{figure}

\begin{table}  
\caption{Calculated lattice parameter $a$, band gap $E_g$, and spin-orbit splitting $\Delta_{\rm SO}$ of AlAs, GaAs, InAs, AlSb, GaSb, InSb, AlBi, GaBi, and InBi, all in zinc blende structure. The negative band gaps of the bismides corresponds to $\Gamma_{6c}$ below the $\Gamma_{8v}$.  Experimental values, from Ref.~\onlinecite{MadelungEd3}, are low-temperature data for overall consistency. AlAs and AlSb have indirect band gaps, while all others have direct band gaps.}
\begin{ruledtabular}
\begin{tabular}{lclrlcl}
 & \multicolumn{2}{c}{$a$(\AA)} 
 & \multicolumn{2}{c}{$E_{\text{g}}$(eV)} 
 & \multicolumn{2}{c}{$\Delta_{\text{SO}}$(eV)} \\
 \cmidrule{2-3}  \cmidrule{4-5}  \cmidrule{6-7}
     & Calc. & Exp.  & Calc.	& Exp.  & Calc.	& Exp.  \\ \midrule
AlAs & 5.677 & 5.661 & 2.234 & 2.23	 & 0.324 & 0.275 \\
GaAs & 5.667 & 5.653 & 1.532 & 1.519	 & 0.374 & 0.346 \\
InAs & 6.108 & 6.058 & 0.401 & 0.418 & 0.397 & 0.38  \\
AlSb & 6.152 & 6.135 & 1.690 & 1.686 & 0.647 & 0.673 \\
GaSb & 6.096 & 6.096 & 0.768 & 0.822	 & 0.710 & 0.756 \\
InSb & 6.519 & 6.479 & 0.238 & 0.258	 & 0.748 & 0.81  \\
AlBi & 6.358 & --	 &-0.449 & 	--	 & 2.055 & 	--    \\
GaBi & 6.322 & --	 &-2.059 & 	--	 & 2.189 & 	--    \\
InBi & 6.716 & --	 & -2.154 & 	--	 & 2.203 & 	--    \\ 
\end{tabular}
\end{ruledtabular}
\label{table2}
\end{table}

\begin{figure}
\centering
\includegraphics[width=0.4\textwidth]{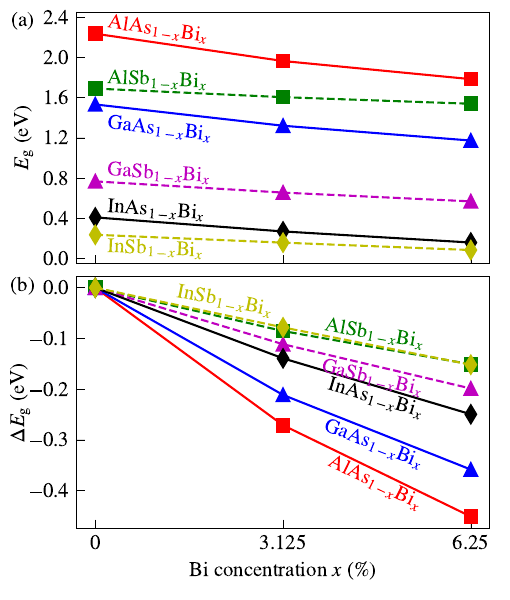}
\caption{Band gap as a function of Bi concentration in dilute III--V$_{1-x}$Bi$_x$ alloys: (a) shows how $E_g$ decreases with increasing Bi percentage. The decrease is larger in the arsenides as compared to the antimonides; (b) shows the relative changes in the band gap, which are mostly linear with Bi concentration. Corresponding values in terms of band-gap reduction per \% of Bi are given in Table \ref{table3}.}
\label{fig3}
\end{figure}

\subsection{Volume variation of dilute bismide III--V alloys with Bi concentration}

Since Bi is a much larger atom than As and Sb (see Table~\ref{table1}), we expect the lattice parameter of the dilute bismide alloys to increase with the Bi content.  The results in Fig.~\ref{fig2}
show that the volume (per 2 atoms) increases almost linearly with Bi concentration, displaying a small deviation from Vegard's law \cite{Vegard} in this dilute concentration regime (more details in Fig.~S3 in the Supplemental Material \cite{SuppMat}. The increase in volume ($\Delta V/V_{\rm 0}$) in the arsenides (Fig.~\ref{fig2}(b)) is visibly larger than that in the antimonides, which is attributed to the much larger difference in atomic radii between As vs Bi compared to Sb vs Bi. The rate of increase is also larger for Ga- and Al- than for the In-based III--Vs. 

\subsection{Band gaps and spin-orbit splittings in the dilute bismide III--V alloys}

The band gaps of the dilute III--V$_{1-x}$Bi$_x$ alloys as a function of Bi content are shown in Fig.~\ref{fig3}(a). As previously reported \cite{Janotti2002,Tixier2003,Polak2015}, we also find a sizable reduction of $E_g$ for such small Bi contents of 3.125\% and 6.25\%, compared to conventional III--V alloys \cite{Wei1990,Vurgaftman2001}. In this low Bi concentration regime, the reduction in the band gap is almost linear as shown in Fig.~\ref{fig3}(b) where $\Delta E_g$ is plotted as a function of Bi concentration. 
The rate of band-gap reduction listed in Table~\ref{table3} fall in the range of the available experimental data and previous calculations for GaAs$_{1-x}$Bi$_{x}$ \cite{Francoeur2003,Huang2005, Kurdawiec2014},  InAs$_{1-x}$Bi$_{x}$ \cite{Okamoto1999,Fang1990, Ma1989,Svensson2012, Mal2022,Zhao2024}, GaSb$_{1-x}$Bi$_{x}$ \cite{Germogenov1989,Das2012,Fluegel2006}, and InSb$_{1-x}$Bi$_{x}$ \cite{Barnett1987,Zilko1980}. It is, however, lower than the value of 90 meV/\%Bi for GaAs$_{1-x}$Bi$_{x}$ reported by Alberi \etal \cite{Alberi2007} for $\sim$1\% Bi.
Note that our values for the band-gap reduction for GaAs$_{1-x}$Bi$_x$ (68 meV/\%Bi) and InAs$_{1-x}$Bi$_x$ (44 meV/\%Bi) are significantly lower than the values of 91 meV\%Bi and 63 meV/\%Bi reported by Polak \etal using the parameterized TB09 MBJLDA functional \cite{Polak2015}.

\begin{table}
\caption{Calculated changes in the valence-band maximum ($\Delta E_v$) and conduction-band minimum $\Delta E_c$, and band gap $\Delta E_g$ in dilute bismide III-(V,Bi) alloys for Bi concentration of 3.125\%.}
\begin{ruledtabular}
\begin{tabular}{lcccc}
 & $\Delta E_v$ & $\Delta E_c$ & \multicolumn{2}{c}{$\Delta E_g$(meV/\%Bi)}  \\
 \cmidrule{4-5}
 & (meV/\%Bi) & (meV/\%Bi) & Calc. & Exp. \& others \\ \midrule 
AlAs$_{1-x}$Bi$_{x}$   & 67 & -20 & 85  & -- \\ 
AlSb$_{1-x}$Bi$_{x}$    & 14 & -13 & 27   & --\\
GaAs$_{1-x}$Bi$_{x}$   & 41 & -27 & 68  & 62--84 \footnotemark[1] \\
GaSb$_{1-x}$Bi$_{x}$   &13 & -22 & 35   & \,\, 32--100 \footnotemark[2]\\
InAs$_{1-x}$Bi$_{x}$   & 29 & -15 & 44  & 20--55 \footnotemark[3]\\
InSb$_{1-x}$Bi$_{x}$   & 11 & -15 & 26   &  17--23 \footnotemark[4]\\
\end{tabular}
\end{ruledtabular}
\footnotetext[1]{Ref.~\cite{Francoeur2003,Huang2005, Kurdawiec2014}}. 
\footnotetext[2]{Ref.~\cite{Germogenov1989,Das2012,Fluegel2006}}
\footnotetext[3]{Ref.~\cite{Okamoto1999,Fang1990, Ma1989,Svensson2012, Mal2022, Zhao2024}}
\footnotetext[4]{Ref.~\cite{Barnett1987,Zilko1980}}
\label{table3}
\end{table}


The largest gap reduction is obtained for the arsenides, with the effect in AlAs$_{1-x}$Bi$_x$ being stronger than in GaAs$_{1-x}$Bi$_x$ and in InAs$_{1-x}$Bi$_x$, in this order. 
This is attributed to the combination of atomic size mismatch between As and Bi vs Sb and Bi and orbital energy differences $\{\epsilon[{\rm Bi}(6p)]-\epsilon[{\rm As}(4p)]\}>\{\epsilon[{\rm Bi}(6p)]-\epsilon[{\rm Sb}(5p)]\}$ (see Table~\ref{table1}).  
In the cases of InAs$_{1-x}$Bi$_x$ and InSb$_{1-x}$Bi$_x$, we predict that the band gap will close for $x\approx0.1$ (10\% Bi), turning these materials into topological insulators. Despite the band gap of InAs being larger than that of InSb, the rate of change in the gap per Bi content is higher in InAs than in InSb. Since InSb$_{1-x}$Bi$_x$ has a lower enthalpy of formation and only about 0.06\% volume change per \%Bi, we expect incorporation of 10\% Bi in InSb to be easier compared to InAs. 

\begin{table}
\caption{Valence-band and conduction-band offsets $\Delta E_v$ and $\Delta E_c$, between dilute III--V$_{1-x}$Bi$_x$ alloys (3.125\% and 6.25\%) and the parent III--V compounds. $\Delta E_g$ is the band-gap reduction and $\Delta[\Delta _{\mathrm{SO}}]$ is increase in spin-orbit splitting. $\Delta E_v$ increases with Bi concentration, as described in the text. The sizable $\Delta E_c$ values are attributed to a deformation potential effect, where the CBM of the III--V$_{1-x}$Bi$_x$ alloys is lowered as the volume increases with Bi concentration.}
\centering
\begin{ruledtabular}
\begin{tabular}{llrrrr}
& $x$ & $\Delta E_v$ & $\Delta E_c$ & $\Delta E_g$ &  $\Delta[\Delta _{\mathrm{SO}}]$\\ 
& (\%) & (meV) & (meV) & (meV) & (meV) \\
\midrule
\multirow{ 2}{*}{AlAs$_{1-x}$Bi$_{x}$} & 3.125 &      208 &     -63  &  271 & 189\\
 & 6.25 &     319 &   -132 &  451 & 282 \\ \midrule
\multirow{ 2}{*}{AlSb$_{1-x}$Bi$_{x}$} & 3.125 &       44   &  -41 & 85 & 57\\
 & 6.25 &    82 &    -69  &  151 & 109\\  \midrule
\multirow{ 2}{*}{GaAs$_{1-x}$Bi$_{x}$} & 3.125 &     127 &    -84 & 211 & 125\\
 & 6.25 &     218 &    -140 & 358 & 212\\ \midrule
\multirow{ 2}{*}{GaSb$_{1-x}$Bi$_{x}$} & 3.125 &       41 &    -70 & 111 & 54\\
 & 6.25 &       72 &   -126 &  198 & 105 \\ \midrule
\multirow{ 2}{*}{InAs$_{1-x}$Bi$_{x}$} & 3.125 &    91   & -48 & 139  & 113\\
 & 6.25 &      162 &   -88 & 250 & 195 \\ \midrule
\multirow{ 2}{*}{InSb$_{1-x}$Bi$_{x}$} & 3.125 &   33   &  -46 & 79 & 53 \\
 & 6.25 &       62 &   -89 &  151 & 102
\end{tabular}
\end{ruledtabular}
\label{table4}
\end{table}

\begin{figure}
\centering
\includegraphics[width=0.4\textwidth]{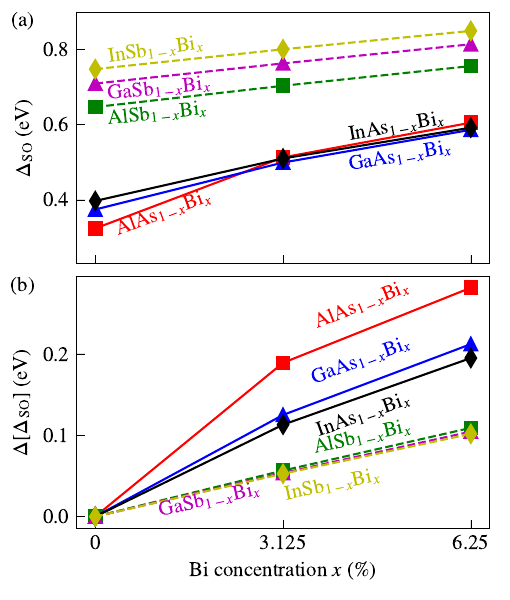}
\caption{(a) Calculated spin-orbit splitting $\Delta_{\rm SO}$ in dilute III--V$_{1-x}$Bi$_x$ alloys, and (b) the rate of change of the spin-orbit splitting with Bi concentration. Except for AlAs$_{1-x}$Bi$_x$ and GaAs$_{1-x}$Bi$_x$, the changes are basically linear in this dilute regime. }
\label{fig4}
\end{figure}

For the spin-orbit splitting at $\Gamma$, we first note that the computed values of $\Delta_{\rm SO}$ for the parent compounds are in good agreement with experimental values \cite{MadelungEd3} and previous calculations \cite{Carrier2004}, as seen in Table~\ref{table2}. In the dilute III--V$_{1-x}$Bi$_x$ alloys, we find that $\Delta_{\rm SO}$ monotonically increases with Bi concentration as shown in Fig.~\ref{fig4}(a). 
As expected, the spin-orbit splittings are systematically larger in the dilute alloys based on Sb than those based on As. 

In the case of AlAs$_{1-x}$Bi$_x$ and GaAs$_{1-x}$Bi$_x$, $\Delta_{\rm SO}$ shows a downward bowing as the Bi content increases. This can be clearly seen in the calculated rate of change of $\Delta_{\rm SO}$ with Bi content, as displayed in Fig.~\ref{fig4}(b). i.e., the rate of change of $\Delta_{\rm SO}$ decreases as the Bi concentration increases from 3.125\% to 6.25\%, and this effect is visibly stronger in AlAs$_{1-x}$Bi$_x$ than in GaAs$_{1-x}$Bi$_x$.  From Fig.~\ref{fig4}(b), we also note that the increase in $\Delta_{\rm SO}$ in the Sb-based alloys is much smaller than in the As-based alloys, and varies linearly with Bi concentration in this dilute range.  It also does not depend on the group-III element. 

Finally, for GaSb$_{1-x}$Bi$_x$, InAs$_{1-x}$Bi$_x$, and InSb$_{1-x}$Bi$_x$ we predict that $\Delta_{\rm SO}$ is larger than the band gap for considered concentrations of 3.125\% and 6.25\%, which indicates that Auger recombination involving a hole in the split-off band will be suppressed.  This is an interesting result for designing lasers that operate in wavelengths longer than 2.1 $\mathrm{ \mu m}$ \cite{Sweeney2014}.

\begin{figure}
\centering
\includegraphics[trim={5mm 2mm 0 12mm},clip, width=0.5\textwidth]{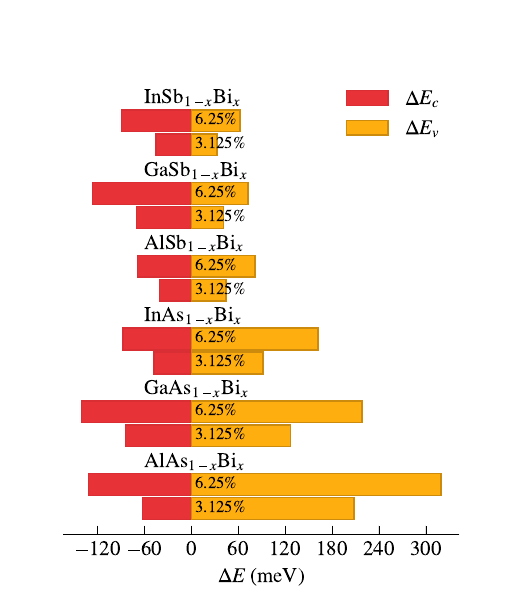}
\caption{Valence-band and conduction-band offsets $\Delta E_v$ and $\Delta E_c$ between the dilute III--(V,Bi) alloys and the corresponding III-V parent compounds, for 3.125\% and 6.25\% Bi content.
The conduction band is also significantly affected. The values in this plot are also listed in Table~\ref{table4}.}
\label{fig5}
\end{figure}

\begin{figure*}
\centering
\includegraphics[width=0.9\textwidth]{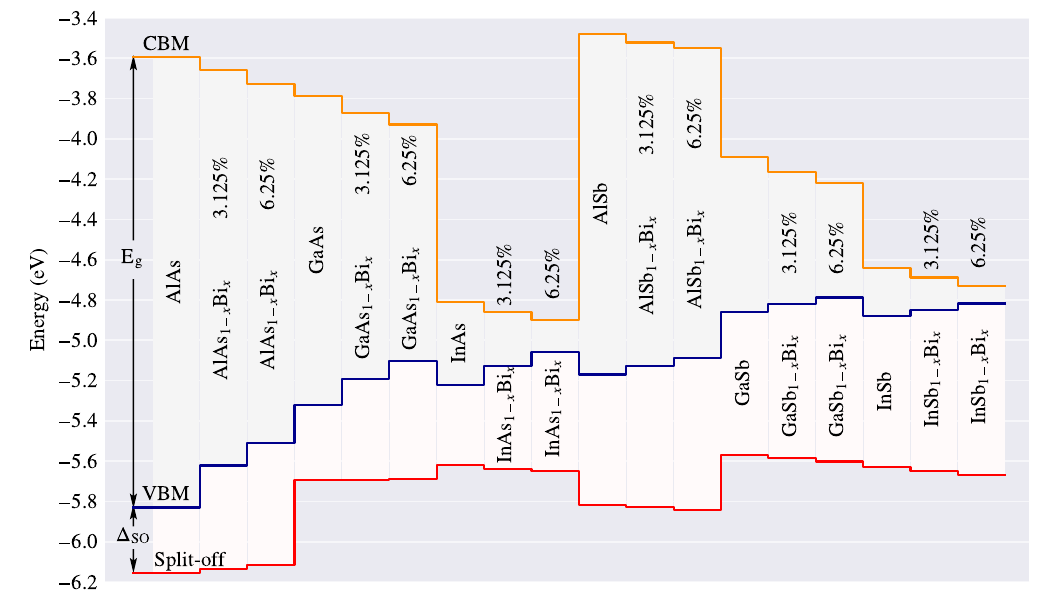}
\caption{Band alignments between the dilute III--V$_{1-x}$Bi$_x$ alloys and the III--V parent compounds.  Values for ionization potential from Ref.~\onlinecite{Andreas} were used to align the VBM of the III--Vs with respect to the vacuum level. As-based alloys show a larger change in VBM and CBM compared to the Sb-based alloys.}
\label{fig6}
\end{figure*}

\subsection{Band alignments between the \texorpdfstring{III--V$_{1-x}$Bi$_x$}{III-V-Bi} alloys and their parent compounds}

The dramatic changes in the band gap of III--V semiconductors by adding a few atomic percent of Bi have often been explained using the VBAC model where the Bi-6$p$-related state, supposedly localized (flat in $k$) and lower in energy than the VBM of the host material, couples with the VBM pushing it up \cite{Alberi2007,Zide2013,Amajdar2014}. The effects of Bi on the CBM are neglected in this simple model.  

To quantify the effects of adding Bi on the valence and conduction bands, we calculated the band alignments between the III--V$_{1-x}$Bi$_x$ alloys and the III-V parent compounds. The magnitude of the shift in the VBM ($\Delta E_v$) and CBM $\Delta E_c$ in III--V$_{1-x}$Bi$_x$ relative to their parent compounds are listed in Table~\ref{table4} and plotted in Fig.~\ref{fig5} for ease of visualization. 
Most notably, the effect of adding Bi on the CBM is as large as that on the VBM for both arsenides and antimonides.
In the case of 3.125\% Bi, the effect on the CBM is significantly larger (in magnitude) than on the VBM in GaSb$_{1-x}$Bi$_x$ (-22 vs 13 meV/\%Bi) and InSb$_{1-x}$Bi$_x$ (-15 vs 11 meV/\%Bi), and basically equal in AlSb$_{1-x}$Bi$_x$ (-13 vs 14 meV/\%Bi). These results are qualitatively different from those calculated using the parameterized TB09 MBJLDA functional \cite{Polak2015} that predict smaller changes in the CBM than in the VBM for 3.7\% Bi concentration.

Adding Bi raises the VBM since the Bi 6$p$ orbital is higher in energy than the As 4$p$ or Sb 5$p$ orbitals that compose the VBM in the III--Vs. 
The decrease in the CBM with the increase in Bi content is attributed to the volume effect. The CBM at $\Gamma$ in the III--Vs are composed mostly of group-III valence $s$ orbitals, i.e., Al 3$s$ in AlAs and AlSb, Ga 4$s$ in GaAs and GaSb, and In 5$s$ in InAs and InSb.  In the dilute III--V$_{1-x}$Bi$_x$ alloys, the CBM is still composed of the same group-III valence $s$ orbitals as in the parent compounds. The absolute deformation potentials for the CBM in the III--Vs are large and negative \cite{Li2006}, meaning that as the volume increases, the CBM rapidly decreases. So in the dilute III--V$_{1-x}$Bi$_x$ alloys, we expect the same effect to occur, i.e., as the volume increases by adding Bi, the CBM decreases. This effect is as large as the contribution from Bi in pushing up the VBM. Note, again, that this effect is not included in the VBAC model by which the CBM would remain unchanged.

In Fig.~\ref{fig6}, we show the band alignments between the dilute III--V$_{1-x}$Bi$_x$ alloys and the parent compounds. The energies are referenced to the vacuum level, and the ionization potentials for the parent III--V compounds are HSE values extracted from Ref.~\onlinecite{Andreas}. The results in Fig.~\ref{fig6}, as in Table~\ref{table4} and Fig.~\ref{fig5}, show stronger effects of Bi in the As-based alloys than in Sb-based alloys. It clearly shows the larger $\Delta_{\rm SO}$ than the band gap in the InAs, GaSb, and InSb-based dilute III--V$_{1-x}$Bi$_x$ alloys. 

We note that as the Bi content increases, the spin-orbit splitting increases; however, the position of the split-off state $\Gamma_{7v}$ with respect to the vacuum level remains practically unchanged, only slightly decreasing in energy as the Bi content increases. We attribute this effect to the strong non-bonding character of the $\Gamma_{7v}$ state, composed mostly of $p_z$ with weak coupling to the heavy hole and light hole bands near $\Gamma$.

\begin{figure}
\centering
\includegraphics[trim={8mm 0 0 0},clip,width=0.5\textwidth]{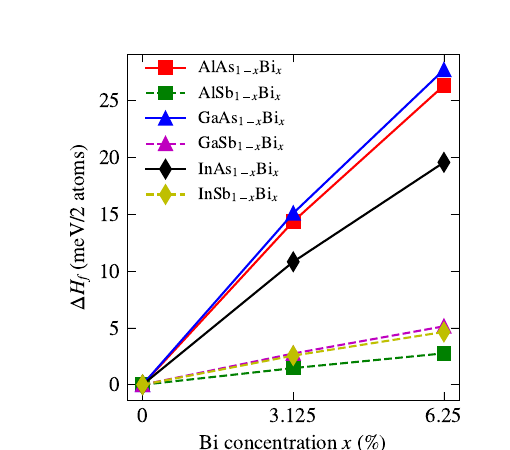}
\caption{Formation enthalpy of dilute III--V$_{1-x}$Bi$_x$ alloys (in meV/2 atoms) as function of Bi concentration. Alloys based on the antimonides have low formation enthalpies because they have a smaller lattice match to III--Bi than the alloys based on the arsenides.}
\label{fig7}
\end{figure}

\subsection{Formation enthalpy of dilute III--V bismide alloys}

Finally, the formability of the dilute III--V$_{1-x}$Bi$_x$ alloys was investigated by computing the formation enthalpy ($\Delta H_f$) of the III--V$_{1-x}$Bi$_x$ as a function of Bi concentration $x$, defined as follows: 
\begin{equation}
\label{en_eq}
\Delta H_f (x)= \mathrm{E}_{\mathrm{III-V}_{1-x}\mathrm{Bi}_x}-(1-x)  \mathrm{E}_{\mathrm{III-V}} - x  \mathrm{E}_{\mathrm{III-Bi}},
\end{equation}
Here, $\mathrm{E}_{\mathrm{III-V}_{1-x}\mathrm{Bi}_x}$ is the total energy of the III--V$_{1-x}$Bi$_x$ alloy, $\mathrm{E}_{\mathrm{III-V}}$ is the total energy of the III--V parent compound, and $\mathrm{E}_{\mathrm{III-Bi}}$ is the total energy of the phase of the III--Bi compound in the metastable zinc blende. 

The calculated $\Delta H_f$ for the III--V$_{1-x}$Bi$_x$ alloys as a function of Bi concentration, shown in Fig.~\ref{fig7}, are higher than those of conventional III--V semiconductor alloys \cite{Wei1988, Seongbok1989}. We note that the antimonides have almost 10 times lower formation enthalpies than the arsenides, indicating that the former are likely to show higher solubility of Bi. This is also attributed to the larger chemical and atomic size mismatch of As vs Bi compared to Sb vs Bi.

\section{Summary}

Using hybrid functional we investigated the the electronic properties of dilute III--V$_{1-x}$Bi$_x$ alloys (3.125\% and 6.25\%), and compared to previous theoretical and experimental studies. We find that both VBM and CBM are affected by the incorporation of Bi, contrary to the simple valence band anticrossing model that assumes that Bi only changes the VBM. The volume increases linearly with Bi concentration and shows only a slight deviation from Vegard's law.  The effects of adding Bi on VBM, CBM, and spin-orbit splitting $\Delta_{\rm SO}$ are systematically larger in the As- than in the Sb-based alloys. At 3.125\% Bi, we predict that the band gap in the In-based alloys will be smaller than $\Delta_{\rm SO}$, suppressing Auger recombination involving holes in the split-off band. We also predict that adding $\approx10\%$ Bi to InAs and InSb will turn them into topological insulators with an inverted band gaps.  The calculated band gaps, spin-orbit splittings, and band offsets serve as a guide to the design of novel electronic and optoelectronic materials based on these large mismatch dilute alloys.

\section{Acknowledgements}

The authors acknowledge fruitful discussions with J. M. O. Zide. This work was supported by the NSF through the UD-CHARM University of Delaware Materials Research Science and Engineering Center (MRSEC) Grant No.~DMR-2011824. S.K. acknowledges funding from the LDRD Program (Project No. 800025) at Princeton Plasma Physics Laboratory under U.S. Department of Energy Prime Contract No.~DE-AC02-09CH11466. We also acknowledge the use of computational resources from the National Energy Research Scientific Computing Center (NERSC), a Department of Energy Office of Science User Facility, through the NERSC award BES-ERCAP 0034471 (m5002), and the DARWIN computing system at the University of Delaware, which is supported by the NSF Grant No.~1919839.

\bibliography{bismides}
\end{document}